\documentclass[secnumarabic,amssymb, amsmath, 12pt,
nofootinbib,tightenlines, nobibnotes, aps, prl]{revtex4}
\usepackage{graphicx}
\usepackage{pstricks}
\begin{document}
\renewcommand{\baselinestretch}{1.3}
\newcommand{\be}{\begin{equation}}
\newcommand{\ee}{\end{equation}}
\newcommand{\ba}{\begin{eqnarray}}
\newcommand{\ea}{\end{eqnarray}}
\hoffset=0.0cm
\voffset=-1.0cm
\textheight=20.0cm
\textwidth=15.5cm

\bigskip

\bigskip

\vspace{2cm}
\title{Tensor mesons produced in tau lepton decays}
\vskip 6ex
\author{G. L\'opez Castro}
\email{glopez@fis.cinvestav.mx}
\affiliation{Departamento de F\'\i sica, Cinvestav, Apdo. Postal
14-740, 07000 M\'exico, D.F. M\'exico}
\author{J. H. Mu\~noz}
\email{jhmunoz@ut.edu.co}
\affiliation{Departamento de F\'\i sica, Universidad del Tolima, A.A.
546, Ibagu\'e, Colombia\\
Centro Brasileiro de Pesquisas Fisicas, Rua Xavier Sigaud 150, 22290-180, Rio de Janeiro, RJ, Brazil}

\bigskip

\bigskip

\begin{center}
\begin{abstract}
 Light tensor mesons ($T=a_2, \ f_2$ and $K_2^*$) can be produced in decays of $\tau$ leptons. In this paper we compute the branching ratios of $\tau \rightarrow T\pi \nu $ decays by assuming the dominance of intermediate virtual states to model the form factors involved in the relevant hadronic matrix elements. The exclusive $f_2(1270)\pi^-$ decay mode turns out to have the largest branching ratio, of $O(10^{-4})$ . Our results indicate that the contribution of tensor meson intermediate states to the three-pseudoscalar channels of $\tau$ decays are rather small.
\end{abstract}
\end{center}

\maketitle
\bigskip

\section{1. Introduction}

Tau leptons are heavy enough that their decay products can contain an on-shell spin-2 tensor meson\footnote{Hereafter, $P$, $V$, $A$ and $T$ will denote the lowest lying pseudoscalar, vector, axial and tensor mesons, respectively.} ($J^{PC}=2^{++}$, see \cite{pdg}) in the final state. Therefore, the $\tau \to TP\nu_{\tau}$ decays can provide a unique environment  to study the weak-tensor-pseudoscalar vertex in the moderate energy regime. Measurements of these hadronic matrix elements will be complementary to the ones involved in the crossed related  $P \to T$  weak transitions which are accessible only in the decays of heavy  $D$ and $B$ mesons.  The hadronic matrix elements $\langle T|J_{\mu}|B\rangle $, which are important in the calculation of semileptonic $B \to Tl\nu$ and non-leptonic $B\to PT, VT$ or $AT$ decays, have been calculated in the framework of several effective models of QCD \cite{isgw,charles,ebert,cheng1,yang,wang,cheng,others,Cheng-Chiang2010}. Manifestly, semileptonic decays provide a cleaner environment to study the weak $PT$ vertex owing to an exact factorization of the decay amplitude, while the non-leptonic amplitudes receive contributions from different terms of the effective weak hamiltonian and a factorization approximation is usually assumed in some calculations (for an extensive literature on the subject see  \cite{isgw,charles,ebert,cheng1,yang,wang,cheng,others,Cheng-Chiang2010,verma,lopez,kim}).  From the experimental side, a few measurements or upper limits about some of these $B$ meson decay channels have been reported so far by $B$-factory experiments (results are listed in \cite{pdg}) and a proper account of the measured rates is still the subject of current investigations. Conversely, tensor mesons produced in $\tau$ lepton decays have been scarcely investigated at the theoretical level and, from the experimental point of view, only the upper limit $B(\tau \rightarrow K_{2}^*(1430)\nu _{\tau })<
3\times 10^{-3}$ has been reported in \cite{pdg}. As it was discussed in \cite{3}, if it was observed, this decay mode would require  the existence of exotic tensor charged weak currents.
 
As is well known, measurements of  $\tau $ decays involving
two or more pseudoscalar mesons have shown the presence of several intermediate
resonant states which populate the different hadronic invariant-mass spectra \cite{Davier:2005xq, Lee:2010tc, Aubert:2007mh}. Indeed, these hadronic spectra have been useful to determine the properties of  $\rho,\ \rho',\ a_1(1260)$ and $K^*$ resonances in a clean environment (see discussion in \cite{Davier:2005xq}). Recently, both BaBar and Belle collaborations have reported refined measurements of $\tau$ decays into three pseudoscalar mesons which include either pions and/or kaons \cite{Lee:2010tc, Aubert:2007mh}. Since tensor mesons undergo sizable decay rates to two pseudoscalar mesons in a $d$-wave orbital configuration \cite{pdg}, one may expect that $T$ mesons give a contribution to three-pseudoscalar $\tau$ lepton decays via the $\tau \to P_1T^*(\to P_2P_3)\nu_{\tau}$ decay chain; eventually, we would be able to extract the $\tau \to TP\nu_{\tau}$ rates from the relevant hadronic spectra as it was done  recently to extract the branching fractions for the  $\tau^- \to \phi \pi^-\nu, \phi K^-\nu, KK^*\nu$ \cite{Lee:2010tc, Aubert:2007mh} decay modes from data on the three-pseudoscalar channels of tau decays.

In this paper we study the $TP$ channels that are kinematically allowed in $\tau$  lepton decays. Most of the popular effective models of QCD at low energies do not make predictions for the weak $PT$ vertex in the energy region relevant for $\tau$ decays. Here we use a meson dominance model where the weak and strong coupling constants are determined from other independent decay processes (see for example \cite{Gabriel2008} for an application to $\tau \to (\omega, \phi)\pi \nu$ decays). We find that the branching fractions for the $\tau \to T\pi\nu$ channels under study are of the order of $10^{-6}\sim 10^{-4}$ and therefore, intermediate tensor resonances give a small contribution to the rates of three-pseudoscalar final states. Eventually, the large data sample of $\tau^+\tau^-$ pairs accumulated by $B$-factory experiments would allow to extract the rates of tensor mesons produced in $\tau$ lepton decays.

%  On the  other hand, the cleanest decay mode $B\rightarrow Tl\nu $
%has not been  observed yet despite that its branching
%fraction is predicted to be of the order of $10^{-3}$ for $B\rightarrow
%D_2^{*}l\nu$ and $10^{-5}\sim 10^{-6}$ for $B\rightarrow a_2l\nu$
%\cite{5}.

\bigskip

\section{2. A meson dominance model for $\protect\tau \rightarrow
T \pi \nu $ decays.}

\bigskip

Let us consider the $\tau^-
(p_{1})\rightarrow T(p_{T})\pi (p)\nu_{\tau} (p_{2})$ decay, where $T$ denotes an on-shell tensor meson; analogous decays involving a kaon or $\eta$ meson are (almost) forbidden by kinematics. The decay amplitude for this process is given by:
\begin{equation}
{\cal M}(\tau \rightarrow T\pi \nu
)=\frac{G_{F}}{\sqrt{2}}V_{uD}\bar{u}%
(p_{2},s_{2})\gamma _{\mu }(1-\gamma _{5})u(p_{1},s_{1})\langle T(p_{T})\pi (p)|J^{\mu }(0)|0\rangle ,  \label{1}
\end{equation}
where $V_{uD}$  ($D=d$ or $s$) is the $uD$ entry of the Cabibbo-Kobayashi-Maskawa matrix and $J_{\mu}(0)$ is the corresponding $V-A$ weak current.

The hadronic matrix element $\langle T\pi |J^{\mu }|0\rangle$
can be parametrized as follows \cite{isgw}:
 \begin{eqnarray}
\langle T(p_{T},\varepsilon )\pi (p)|J^{\mu }|0\rangle
&=&ih\epsilon ^{\mu \nu \phi \rho }\varepsilon _{\nu \alpha }^{\ast
}p^{\alpha }(p-p_{T})_{\phi }(p+p_{T})_{\rho }-k\varepsilon ^{\ast \mu \nu
}p_{\nu }  \nonumber \\
&&\ -\varepsilon ^{\ast \alpha \beta }p_{\alpha }p_{\beta }[b_{+}(p-p_{T})^{\mu
}+b_{-}(p+p_{T})^{\mu }],  \label{2}
\end{eqnarray}
where $h(t),$ $k(t)$, $b_{\pm }(t)$ are  Lorentz-invariant form factors and $t=(p_T+p)^{2}$ is the square of the momentum transfer. The symmetric tensor $\varepsilon^{\ast \mu\nu}$ describes the spin-2 polarization states of the  outgoing tensor meson.

The unpolarized squared amplitude becomes:

\begin{equation}
\sum_{pols}|{\cal M}|^{2}=2G_{F}^{2}|V_{uD}|^{2}\left[
c_{1}(u,t)|h|^{2}+c_{2}(u,t)|k|^{2}+c_{3}(u,t)|b_{-}|^{2}+c_{4}(u,t) {\rm
Re}(k^{\ast }b_{-})\right]\ ,  \label{3}
\end{equation}
where we have defined $u=(p_1-p)^2$, and  $c_i=c_i(u,t)$ are kinematical
factors given by:
\begin{eqnarray}
c_1&=&\frac{\lambda }{8m_{T}^{2}}\left\{2tu^{2}+2\left[ m_{\tau
}^{2}(m_{\pi
}^{2}-m_{T}^{2})-t(m_{T}^{2}+m_{\tau }^{2}+m_{\pi }^{2}-t)\right] u
\right. \nonumber \\ &&  \left. +\frac{1}{2}(m_{\tau }^{2}-t) \left[
-2\lambda  +t(2m_{\pi
}^{2}+2m_{T}^{2}-t)-m_{\tau }^{2}(m_{\tau }^{2}-t)-6m_{T}^{2}m_{\pi
}^{2}\right] \right. \nonumber \\
&&  \left. +\frac{1}{2}m_{\tau  }^{2}(2m_{T}^{2}+m_{\tau
}^{2}-t)^{2}\right\},
\\ \label{4}
c_{2}&=& \frac{1}{12m_{T}^{4}}\left\{(\lambda +tm_{T}^{2})u^{2}+\left[
(t-m_{\pi}^{2}-m_{T}^{2})(\lambda +m_{T}^{2}t)-m_{T}^{2}m_{\tau
}^{2}(t+m_{T}^{2}-m_{\pi }^{2})\right]u \right. \nonumber \\
&& \left. +\frac{1}{2}m_{T}^{2}(m_{\tau }^{2}+2m_{\pi }^{2}-3t)\lambda
+m_{T}^{4}m_{\tau }^{2}(m_{\tau }^{2}+m_{T}^{2}-t)-m_{\pi
}^{2}m_{T}^{4}(m_{\tau }^{2}-t)\right\}, \\ \label{5}
c_{3}&=& \frac{m_{\tau }^{2}(m_{\tau }^{2}-t)}{48m_{T}^{4}}\lambda ^{2},
\\ \label{6}
c_{4}&=& \frac{m_{\tau }^{2}\lambda
}{12m_{T}^{4}}\left\{(t+m_{T}^{2}-m_{\pi
}^{2})u-m_{T}^{2}(2m_{\tau }^{2}+m_{T}^{2}-m_{\pi }^{2}-t)\right\}.
\label{7}
\end{eqnarray}
where $\lambda = t^2 + m_T^4+m_{\pi}^4-2tm_{T}^2-2tm_{\pi}^2-2m_T^2m_{\pi}^2$.

\begin{figure}
  \includegraphics[width=18cm]{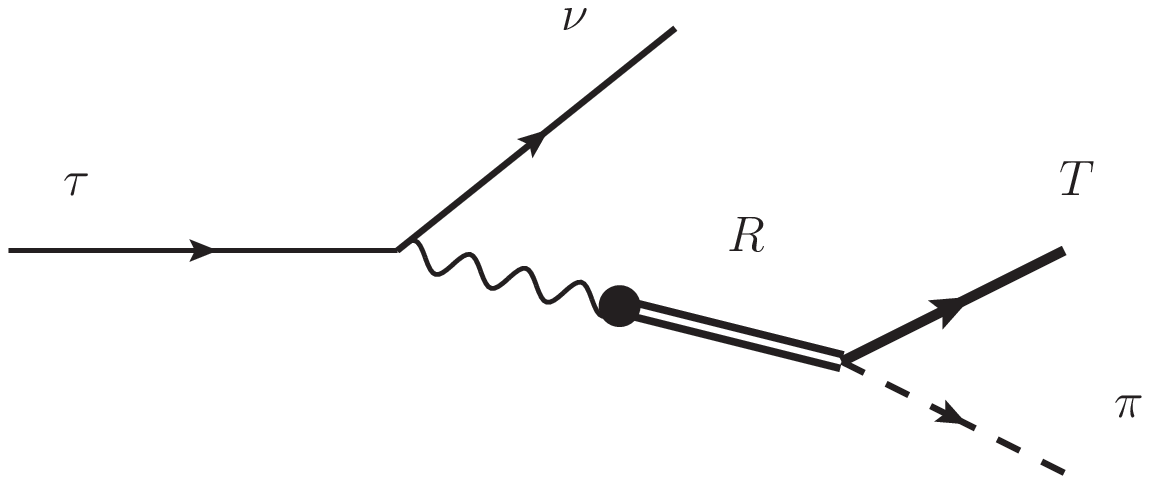}
\vspace{-18.0cm}
  \caption{Intermediate meson dominance graph of $\tau \to T\pi\nu$ decays}\label{fig1}
\end{figure}

As we have pointed out previously, we resort to a meson dominance model to compute the form factors in our $\tau$ decays (see Figure \ref{fig1}). For definiteness, we will illustrate the method in the case of the $\tau \rightarrow K_{2}^{\ast }(1430)\pi \nu $ decay, because in this case all form factors receive contributions from intermediate $t$-channel virtual states. 
In this model we will assume that the above decay receives contributions from three intermediate states: the pseudoscalar $K$ and axial $K_{1}(1400)$ mesons which saturate the axial current, and the vector meson $K^{\ast }(892)$ which contributes to the vector current. Other meson resonances can
contribute as well to both currents; we would expect their corrections to be small since either their strong couplings to the $K_2^*\pi$ system or their couplings to the weak current are suppressed. Such additional contributions may be enhanced if their resonance shapes were peaked in the kinematical domain of $\tau$ decays ($(m_T+m_{\pi})^2 \leq t \leq m_{\tau}^2$), which is not the case. 

Within our approximations, the decay amplitude is given by (see Figure 1)
\begin{equation}
{\cal M} (\tau \rightarrow K_{2}^{\ast }\pi \nu )=\sum_{R=K, \ K^*(892), \ K_1(1400) }\!\!\!\!\!\!\!\!\!\!{\cal M}(\tau
\rightarrow  R\nu
\rightarrow K_{2}^{\ast }\pi \nu )\ .
\label{8}
\end{equation}

%%%%%%
%%%%%%%%%%%%

  Using the Feynman rules to compute the above amplitudes and comparing the results with Eq. (\ref{2}), we derive the following expressions
for the form factors:
\begin{eqnarray}
k(t)&=&- \frac{if_{K_{1}}g_{K_{2}^{\ast }K_{1}\pi
}}{m_{K_1}}\cdot {\rm BW}_{K_1}(t),  \\ \label{9}
h(t)&=&\frac{f_{K^{\ast }}g_{K_{2}^{\ast }K^{\ast }\pi
}}{2m_{K^*}}\cdot {\rm BW}_{K^*}(t),
\\ \label{10}
b_{-}(t)&=&\frac{if_{K_{1}}g_{K_{2}^{\ast }K_{1}\pi }}{m_{K_{1}}^3}\cdot {\rm BW}_{K_1}(t)+ 
\frac{f_{K}g_{K_{2}^{\ast }K\pi }}{m_K^2}\cdot {\rm BW}_K(t), \\
\label{11} b_{+}(t)&=& 0,  \label{12}
\end{eqnarray}
where $f_X$ and $g_{K_2^* X\pi}$ denote the weak and strong coupling constants of
intermediate $X$ states. The Breit-Wigner forms introduced above are defined as ${\rm BW}_X(t)=m_X^2/(m_X^2-t-im_X\Gamma_X \theta(t-t_{\rm threshold}))$, with $m_X$ and $\Gamma_X$ being the mass and decay width (which we choose to be a constant) of the resonance. 

In a similar way, we can assume the same meson dominance model to describe the 
strangeness-conserving  $\tau \rightarrow a_{2}(1320)\pi \nu$  decay.
Owing to G-parity conservation \cite{scc}, the amplitude for the $\tau \rightarrow a_{2}\pi \nu$ process will receive 
contributions only in the vector current via the following decay chain $\tau \to
(\rho(770), \rho'(1450))\nu \to a_2(1320)\pi\nu$, where $\rho'$ is the first radial excitation of the $\rho$ meson. In this case, the only  non-vanishing form  factor  becomes:
\begin{equation}
h(t)=\ \frac{f_{\rho }g_{a_{2}\rho \pi }}{2m_{\rho}}\cdot  \frac{{\rm BW}_{\rho}(t)+\beta {\rm BW}_{\rho'}(t) }{1+\beta}\ ,  \label{13}
\end{equation}
where $\beta$ denotes the ratio of $\rho'$ to $\rho$ coupling constants, and it is similar to one defined in the two-pseudoscalar decay modes (see for example \cite{beta}). In the studies of these two-pseudoscalar decays of $\tau$ leptons  carried out by ALEPH \cite{taupipi} and Belle \cite{taukpi} Collaborations, $\beta$ turns out to be small and almost real: $\beta_{\pi\pi} \approx -0.15 $ \cite{taupipi} and $\beta_{K\pi} \approx 0.08$ \cite{taukpi}, for the $\pi^-\pi^0$ and $K^0\pi^-$ decay modes, respectively. In our present calculation, we will assume $\beta_{a_2\pi}=(\pm 0.2 \pm 0.1)$ as a rather conservative value. 

Finally, we also consider  the  $\tau \rightarrow f_{2}(1270)\pi \nu$ decay. In this case $G$-parity conservation forbids the contribution of the vector current to the decay amplitude. We assume that  the dominant contributions come from the pseudoscalar and axial resonances by means of the chain $\tau \to (\pi,\ a_1) \nu \to f_2 \pi \nu$. We also assume that the mixing angle between the $f_2(1270)$ and $f_2'(1525)$ tensor mesons is such that the $f_2$ is dominantly a $(u\overline{u} + d\overline{d})/\sqrt{2}$ state \cite{Cheng-Chiang2010}. The only non-vanishing form factors in this case become:
\begin{eqnarray}
k(t)&=&- \frac{if_{a_{1}}g_{f_{2}a_{1}\pi
}}{m_{a_1}}\cdot {\rm BW}_{a_1}(t),  \\ \label{14}
b_{-}(t)&=&\frac{if_{a_{1}}g_{f_{2}a_{1}\pi }}{m_{a_{1}}^3}\cdot {\rm BW}_{a_1}(t)+ 
\frac{f_{\pi}g_{f_{2}\pi\pi }}{m_{\pi}^2}\cdot {\rm BW}_{\pi}(t), \label{15}
\end{eqnarray}

\bigskip

\section{3. Determination of the strong and weak couplings.}

\bigskip

In this section we focus on the determination of the strong and weak coupling
constants that appear in Eqs.(9-15). We first consider in more detail
the decay widths of the  $T\rightarrow P\pi,\ V\pi, \ A\pi$,
and $A\rightarrow V\pi$ decays reported in Ref. \cite{pdg} to determine the strong couplings.

The decay constants for the above decays are defined from the following decay amplitudes \cite{7}, which assume that only one single $L$-wave configuration contributes to the final states:
\begin{eqnarray}
{\cal M}[T(p_{T},\varepsilon )\longrightarrow P(p_{P})\pi (p)]&=&
g_{TP\pi}\varepsilon ^{\mu \alpha }p_{\mu }p_{\alpha } \\ \label{14}
{\cal M} [T(p_{T},\varepsilon )\longrightarrow V(\epsilon ,p_{V})\pi
(p)]&=& g_{TV\pi }\epsilon _{\mu \nu \rho \sigma }\epsilon ^{\mu
}p_{V}^{\nu }\varepsilon^{\rho \alpha }p_{\alpha }p^{\sigma } \\
\label{15}
{\cal M}[T(p_{T},\varepsilon )\longrightarrow A(\epsilon ,p_{A})\pi
(p)]&=& g_{TA\pi }\varepsilon _{\alpha \beta }\epsilon ^{\alpha }p^{\beta
} \\ \label{16}
{\cal M}[A(\epsilon _{A},p_{A})\longrightarrow V(\epsilon
_{V},p_{V})P(p)]&=& g_{AVP}(\epsilon _{A})_{\mu }(\epsilon _{V})_{\nu
}\left\{ p_{A}.p_{V}g^{\mu \nu }-p_{A}^{\nu }p_{V}^{\mu }\right\} .
\label{17}
\end{eqnarray}
The corresponding decay rates in the rest frame of the decaying particle are:
\begin{eqnarray}
\Gamma (T\longrightarrow P\pi )&=&\frac{g_{TP\pi }^{2}}{60\pi m_{T}^{2}}%
\left\vert \vec{P}_{c}\right\vert ^{5}\ , \\ \label{18}
\Gamma (T\longrightarrow V\pi )&=&\frac{g_{TV\pi }^{2}}{40\pi }\left\vert
\vec{P}_{c}\right\vert ^{5}\ , \\ \label{19}
\Gamma (T\longrightarrow A\pi )&=& \frac{g_{TA\pi }^{2}}{120\pi m_{T}^{2}m_{A}^{2}}\left( 2\left\vert
\vec{P}_{c}\right\vert ^{5}+5m_{A}^{2}\left\vert \vec{P}_{c}
\right\vert^{3}\right) \ , \\  \label{20}
\Gamma (A \rightarrow VP)&=& \frac{g_{AVP}^{2}}{8\pi m_{A}^{2}m_{V}^{2}}
\left[m_{A}^{2}|\vec{P}_{c}|^{5}-\frac{1}{4}(m_{A}^{2}+m_{V}^{2}
-m_{P}^{2})^{2}| \vec{P}_{c}|^{3}\right.  \nonumber \\
&& \left. +\frac{3}{4}m_{V}^{2}(m_{A}^{2}+m_{V}^{2}-m_{P}^{2})^{2}|
\vec{P}_{c}|\right]\ ,
\end{eqnarray}
where $\vec{P}_c$ denotes the three-momentum of anyone of the particles in the final state.

In order to extract the decay constant $g_{K_{2}^{\ast }K_{1}\pi }$ we
have assumed that the experimentally measured rate of $K_{2}^{\ast
}(p_{T})\rightarrow  K^{\ast }(p_{V})\pi (p_{2})\pi (p_{1})$ is
saturated by the  contribution of the $K_1(1400)$ intermediate state
through the chain process  $K_{2}^{\ast }\rightarrow K_{1}(1400)\pi
\rightarrow  K^{\ast }\pi \pi $. The dominance of this mechanism is also assumed in other works (see for example \cite{axial}). Of course,  we are aware that other intermediate resonances might also contribute (for example the $\rho$, $\sigma$ and $f_0$ resonances in the $\pi\pi$ channel), but either their couplings to
$K_2^*K^*$ are small or forbidden (an alternative view of the problem is discussed in Ref. \cite{singer}, which considers that the dominant contribution arises from the $g_{TVV}$ coupling).

We assume isospin symmetry to relate the strong coupling constants for different charge states in a given channel,  and SU(3) flavor symmetry to relate the couplings of vertices that can not be measured directly to the ones that are extracted from measured rates. Using these approximations and  the measured rates \cite{pdg} of relevant decays, we obtain the following central values:
$g_{K_{2}^{\ast -}K^{-}\pi ^{0}}=8.35$ GeV$^{-1}$, $g_{K_{2}^{\ast
-}K^{\ast -}\pi ^{0}}=9.02$ GeV$^{-2},$ $g_{K_{1}^{-}K^{\ast -}\pi
^{0}}=1.95$ GeV$^{-1}$ , $g_{K_{2}^{\ast -}K_{1}^{-}\pi ^{0}}=30.6$, 
$\ g_{\overline{K_{2}^{\ast 0}}K_{1}^{-}\pi^+}=-43.3$, $g_{f_2a_1^+\pi^-}=32.3$, 
$g_{\overline{K_{2}^{\ast 0}}K^{-}\pi^+}=12.94$ $\ $GeV$^{-1},\ g_{\overline{K_{2}^{\ast 0}}K^{\ast -}\pi ^{+}}=12.26$ GeV$^{-2},\
g_{a_{2}^{-}\rho ^{-}\pi ^{0}}=19.40$ GeV$^{-2}$, $g_{a_{2}^{0}\rho
^{-}\pi ^{+}}=19.47$ GeV$^{-2}$, and  $g_{f_2\pi^+\pi^-}=20.3$ GeV$^{-1}$. In order to get some of the couplings involving the strange axial mesons \cite{Gabriel2008} we have assumed a value of $\theta_A=50.8^{\circ}$ \cite{Cheng-Chiang2010} for the mixing angle of the $K_1(1270)$ and $K_1(1400)$ strange mesons. We also note that the value of the $f_2\pi\pi$ coupling given above agrees with the prediction obtained in the Appendix of the first paper in Ref. \cite{singer}. We note that the uncertainties associated to these couplings are estimated directly from their measured masses and rates or, when data was not available, they were attributed a conservative $20\%$ uncertainty if SU(3) symmetry was assumed in their derivation.

Finally, the values of other relevant inputs to determine the branching ratios
of tau decays have been taken from Ref. \cite{pdg}. In addition we have
set the weak coupling of hadron $H^-$ from $\tau^- \to
H^-\nu$ decays: $f_{\pi}=(130.7\pm 0.4)$ MeV,  $f_{K^{-}}=(159.8\pm 1.5)$ MeV, $f_{K^{\ast -}}=(210\pm 5)$ MeV, and
$f_{\rho }=(218\pm 2)$ MeV.  On the other hand, we use $f_{a_1}= (238\pm 10)$ MeV in our calculations and we have taken $f_{K_1(1400)}= (-139^{+41}_{-46})$ MeV from Ref. \cite{Cheng-Chiang2010}.

\section{4. Branching ratios of tau decays into tensor mesons.}

 The branching ratios predicted in this work for the $\tau^- \to T\pi\nu$ decays are shown in Table 1. The main uncertainty in the rates of Cabibbo-suppressed channels comes from the large error bar in the $f_{K_1}$ coupling constant, while the one in the $a_2\pi$ channel is dominated by the large uncertainty ($\pm 50\%$) that we have attributed to the value of $\beta$. Since our model uses Breit-Wigner forms with a constant decay width and given that the contributions of higher mass virtual states have been neglected in our calculation of the form factors, further uncertainties are expected to contribute to the results shown in Table 1. 
 In addition, in our model we have not considered the contribution of the continuum which can be associated to a contact (non-BW) term in the weak $T\pi$ vertex. We have not estimated these uncertainties in Table 1; eventually, the continuum contributions may be large, but they are rather difficult to evaluate in the meson dominance model like ours in the absence of constraints about such  contact terms.

\begin{table}[t]
%\begin{table}
\begin{tabular}{|c|c|c|}
\hline
\ \ \ $T\pi$ mode & \ \ \ \ \ \ Branching ratio & \ \ \ \ \ \ Comment\ \ \  \\
\hline \hline 
$K_2^{*-}(1430)\pi^0$ & $(3.7\pm 2.1)\times 10^{-6}$ &  \\
\hline
$\overline{K}_2^{*0}(1430)\pi^-$ &$(4.7\pm 2.7) \times 10^{-6}$ & \\
\hline
$a_2^-(1320)\pi^0$ & $(9.4\pm 4.8)\times 10^{-6}$  & $\beta=+0.2\pm 0.1$,\\   
& $(10.1\pm 7.5)\times 10^{-6}$ & $\beta=-0.2\pm 0.1$ \\
\hline
$a_2^0(1320)\pi^-$ & $(9.0\pm 4.6)\times 10^{-6}$  & $\beta=+0.2\pm 0.1$,\\   
& $(9.5\pm 7.1)\times 10^{-6}$ & $\beta=-0.2\pm 0.1$ \\
\hline 
$f_2(1270)\pi^-$ & $(5.9\pm 1.8)\times 10^{-4}$ & $f_2$ pure $\bar{u}u,\ \bar{d}d$ \\
\hline
\end{tabular}
\caption{Branching ratios for the $(T\pi)$ decays of the $\tau$ lepton.}
\end{table}

The branching fractions turn out to be of order $10^{-4}\sim 10^{-6}$, with the largest rate corresponding to the $f_2(1270)\pi^-$ decay  mode. Therefore, we can expect that the contribution of tensor meson intermediate states to the three-pseudoscalar decays of tau leptons is small. Concerning the results shown in Table 1, we observe that the Cabibbo-favored decay involving the $f_2$ meson is larger that the one involving the $a_2$ meson because, owing to $G$-parity, the former receives contributions from the dominant axial current  while the second is mediated by the vector current only. Similarly, the Cabibbo-suppressed channels are of similar size as the Cabibbo-allowed $a_2\pi$ decays because the former receive contributions of the vector and axial currents. In addition to the above dynamical considerations, we should point out that these $T\pi$ channels are suppressed mainly due to the reduced phase space available in $\tau$ lepton decays.

\bigskip

\bigskip

\section{5. Conclusions}

We have studied and computed the branching ratios of the $\tau \rightarrow (K_{2}^{\ast }, a_{2}, f_2)\pi \nu $ decays; we have not considered final states involving kaons or $\eta$ mesons because either they are suppressed or forbidden by kinematics. We have used a meson dominance model where the form factors are dominated by the lowest lying resonances that couple to the $T\pi$ system. To our knowledge, this is the first study reporting results on these peculiar $\tau$ lepton decays. Beyond probing the tensor-pseudoscalar weak vertex, the processes under consideration can contribute as intermediate states in $\tau$ lepton decays involving three-pseudoscalar mesons. 

Owing to $G$-parity of strong interactions, the rates of the Cabibbo allowed and suppressed decay channels exhibit an interesting pattern. The Cabibbo-supressed $K_2^*\pi$ channels turn our to be of the same order as the Cabibbo-favored $a_2\pi$ decays, mainly because the latter receives contributions only from the vector current.   The calculated branching fractions spread from $10^{-4}$ to $10^{-6}$, with the largest branching fraction ($\approx 6\times 10^{-4}$) corresponding to the $f_2(1270)\pi^-$ final state. Eventually, these decays will be measured from the invariant mass distributions of the decay products of the intermediate $T$ tensor mesons in three-body decays of $\tau$ leptons, given the large data sample of $\tau$ lepton pairs recorded by the Babar and Belle experiments \cite{9}.

\bigskip

\subsection{Acknowledgements}
GLC acknowledges financial support
from Conacyt and SNI (M\'{e}xico). JHM is grateful to \textit{%
Comit\'{e} Central de Investigaciones} (CCI) of the University of Tolima and CNPq (Brazil) for
financial support and also thanks to the Physics Department at Cinvestav for
the hospitality while this work was ended.

\end{document}